\documentclass[english]{revtex4-1}
\usepackage[LGR,T1]{fontenc}
\usepackage[latin9]{inputenc}
\usepackage{float}
\usepackage{mathrsfs}
\usepackage{bm}
\usepackage{amsmath}
\usepackage{graphicx}
\usepackage{setspace}

\makeatletter

\DeclareRobustCommand{\greektext}{%
  \fontencoding{LGR}\selectfont\def\encodingdefault{LGR}}
\DeclareRobustCommand{\textgreek}[1]{\leavevmode{\greektext #1}}
\ProvideTextCommand{\~}{LGR}[1]{\char126#1}

\usepackage{fleqn}
\usepackage{epsfig}\usepackage{slashed}
\usepackage{bbold}

\def\ps {\not\!p}

\def\qs {\not\!q}

\def\rf#1{{(\ref{#1})}}
\def\br{\begin{eqnarray}}
\def\er{\end{eqnarray}}

\usepackage{babel}

\usepackage{babel}

\makeatother

\usepackage{babel}
\begin{document}

\title{\textsc{\Large{}neutrino-induced single pion production and the reanalyzed
bubble chamber data }}

\author{{\normalsize{}D.F. Tamayo Agudelo$^{1}$,A. Mariano$^{2,3}$, and
$^{1}$D.E. Jaramillo Arango}}

\affiliation{\textrm{$^{1}$Facultad de C. Exactas y Naturales Universidad de Antioquia,
Ciudad Universitaria: Calle 67 N$^{0}$ 53-108 Bloque 6 Oficina 105,
Medellin, Colombia, Instituto de F\'{\i}sica.~~ }\\
\textrm{$^{2}$Facultad de Ciencias Exactas Universidad Nacional de
La Plata,C.C. 67,1900 La Plata,Argentina. ~~}\\
\textrm{ $^{3}$Instituto de F\'{\i}sica La Plata CONICET, diagonal 113
y 63, 1900 La Plata, Argentina. ~~}~\\
 ~\\
}
\begin{abstract}
{\normalsize{}In this work we report the calculation of the charged
current total and differential cross sections for weak pion-production
with neutrinos and antineutrinos, with the final pion-nucleon pair
invariant mass $W_{\pi N}\lesssim2$GeV. Our results are compared
with the recent reanalyzed data from the old bubble chamber experiments,
that solved the discrepancy between the ANL and BNL data. We implement
a model previously tested for the cuts $W_{\pi N}<1.4,1.6$ Gev which
includes explicitly resonances in the first and second resonance regions,
within different approaches for the resonances self energy and }$\frac{3}{2}$-{\normalsize{}vertexes
and propagators, a fact not usually analyzed. Our model leans on consistent
effective Lagrangians that generate resonant amplitudes together non
resonant plus resonant backgrounds. Effects of hadrons finite extension
and more energetic resonances corresponding to the emitted pion-nucleon
invariant mass in the $1.6<W_{\pi N}<2$ GeV region, are taking into
account using appropiate form factors in consistency with previous
results on neutral current pion production calculations. Our results
reproduce well the reanalysed data without cuts and are compared with
another models.}{\normalsize \par}

\smallskip{}
 PACS numbers :13.15.+g,13.75.-n,13.60.Le 
\end{abstract}
\maketitle

\section{INTRODUCTION}

For current and future neutrino oscillation experiments we need to
understand single pion production by neutrinos with few-GeV energies.
The pion production is either a signal process when scattering cross
sections are analyzed, or a large background for analyses which select
quasielastic events. At these energies the dominant production mechanism
is via the excitation and subsequent decay of hadronic resonances.
Experimental data on nuclear targets present a confusing picture,
shown from the MINER\textgreek{n}A \cite{Eberly15,Le15}and MiniBooNE
\cite{Are11} experiments in poor agreement with each other in the
framework of current theoretical models\cite{Sob15,Ulrich15}. Complete
models of neutrino\textendash nucleus single pion production interactions
are usually factorized into three parts: the neutrino\textendash nucleon
cross section; additional nuclear effects which affect the initial
interaction; and the \textquotedblleft final state interactions\textquotedblright{}
(FSI) of hadrons exiting the nucleus. 

More basically at the level of neutrino-nucleon cross section, the
axial form factor(FF) for pion production on free nucleons cannot
be constrained by electron scattering data, used normally to get the
vector FF, so it relies upon data from Argonne National Laboratory\textquoteright s
12 ft bubble chamber (ANL)\cite{Rad82} and Brookhaven National Laboratory\textquoteright s
7 ft bubble chamber (BNL)\cite{Kitagi86}. The ANL neutrino beam was
produced by focusing 12.4 GeV protons onto a beryllium target. Two
magnetic horns were used to focus the positive pions produced by the
primary beam in the direction of the bubble chamber, these secondary
particles decayed to produce a predominantly $\nu_{\mu}$ peaked at
$\sim0.5$ GeV. The BNL neutrino beam was produced by focusing 29
GeV protons on a sapphire target, with a similar two horn design to
focus the secondary particles. The BNL $\nu_{\mu}$ beam had a higher
peak energy of $\sim1.2$ GeV, and was broader than the ANL one. These
both datasets differed in normalization by 30\textendash 40$\%$ for
the leading pion production process $\nu_{\mu}p\rightarrow\mu^{-}p\pi^{+}$,
which conduced to large uncertainties in the predictions for oscillation
experiments as well as in the interpretation of data taken on nuclear
targets\cite{Nieves07,Leitner09,Ahn03,Nieves10,Abe13,Lala13}.

It has long been suspected that the discrepancy between ANL and BNL
was due to an issue with the normalization of the flux prediction
from one or both experiments, and it has been shown by other authors
that their published results are consistent within the experimental
uncertainties provided\cite{Graczyk09,Graczyk14}. In Ref. \cite{wilki14},
was presented a method for removing flux normalization uncertainties
from the ANL and BNL $\nu_{\mu}p\rightarrow\mu^{-}p\pi^{+}$ measurements
by taking ratios with charged-current quasielastic (CCQE) event rates
in which the normalization cancels. Then, it was obtained a measurement
of $\nu_{\mu}p\rightarrow\mu^{-}p\pi^{+}$by multiplying the ratio
by an independent measurement of the charged quasielastic (CCQE) cross
section which is well known for nucleon targets. Using this technique,
they found good agreement between the ANL and BNL $\nu_{\mu}p\rightarrow\mu^{-}p\pi^{+}$datasets.
Later, they extend that method to include the subdominant $\nu_{\mu}n\rightarrow\mu^{-}p\pi^{0}$
and $\nu_{\mu}n\rightarrow\mu^{-}n\pi^{+}$channels\cite{Rodriguez16}.
These authors used the resulting data to fit the parameters of the
GENIE pion production model\cite{Genie10}. They found a set of parameters
that can describe the bubble chamber data better than the GENIE default
ones, and provided updated central values and reduced uncertainties
for use in neutrino oscillation and cross section analyses which use
the GENIE model. In this model the cross section is cutted off at
a tunable invariant mass value, which is $W_{\pi N}\le1.7$ GeV by
default. No in-medium modifications to resonances are considered,
and interferences between resonances are neglected in the calculation
being the Rein\textendash Sehgal(RS) model in Ref.\cite{Rein81} adopted.
Nevertheless, the original RS model includes non-resonant single pion
production as an additional resonance amplitude, while in GENIE the
non-resonant component is implemented as an extension of the deep
inelastic scattering model. They found that GENIE\textquoteright s
non-resonant background prediction has to be significantly reduced
to fit the data, which may help to explain the recent discrepancies
between simulation and data observed by the MINER\textgreek{n}A coherent
pion and NO\textgreek{n}A\cite{Adanson2016} oscillation analyses.While
more sophisticated single pion production models exist, the GENIE
generator is widely used by current and planned neutrino oscillation
experiments, so tuning the generator parameters represents a pragmatic
approach to improving its description of available data. We find that
the reanalyzed data, where the normalization discrepancy has been
resolved, is able to significantly reduce the uncertainties on the
pion production parameters. 

This is one of the reasons encourage us to return to the calculation
of neutrino-nucleon cross sections in our model and try to extend
it to larger final $W_{\pi N}$ invariant masses. On the other hand,
there are many models to describe this process that do not fulfill
several important ingredients:\\
i)There are problems from the formal point of view. The main pion
emision source are excitation and decay of resonances, and many of
them are of spin $\frac{3}{2}$ where its field is built as $\Psi_{\mu}\equiv\psi\otimes\xi_{\mu}$,
where $\psi$ is a Dirac spinor field and $\xi_{\mu}$ is a Dirac
4-vector \cite{Kirbach2002}. In this way, the field $\Psi_{\mu}$
will contain a physical spin-$\frac{3}{2}$ sector and a spurious
spin-$\frac{1}{2}$ sector dragged by construction. Nevertheless,
involved Lagrangians must lead to amplitudes invariant by contact
transformations\cite{Badagnani12}$\Psi^{\nu}\rightarrow\Psi^{'\nu}=R_{\mu\nu}(\frac{1+3A}{2})\Psi^{\nu},R^{\rho\sigma}(a)=g^{\rho\sigma}+a\gamma^{\rho}\gamma^{\sigma}$,
which change the amount of the spin-$\frac{1}{2}$ espurious contribution
since there exist the constraint $\Psi_{\mu}\gamma^{\mu}=0$ for the
$\frac{3}{2}$ sector. Many works keep the simpler forms of both the
free and interaction Lagrangians involving $\Psi_{\mu}$, that correspond
to different $A$ values, and with this choice amplitudes lacks the
mentioned invariance. In addition, the interaction Lagrangian for
the spin-$\frac{3}{2}$ field coupled to a nucleon N ($\psi$) and
a pseudo-scalar meson ($\phi$) or boson ($W$), as usually appears
in a resonance production-decay, depends on a second parameter $Z$
not fixed by the contact invariance. Now to fix it, we point to the
question of the true degree of freedom of the spin-$\frac{3}{2}$
field, which dynamics is described by a constrained quantum field
theory. Observe that in the free $\Psi_{\mu}$ Lagrangian there is
no term containing $\dot{\Psi}_{0}$ \cite{Badga17}. So, the equation
of motion for it is a true constraint, and $\Psi_{0}$ should have
no dynamics. It is necessary then that interactions do not change
that fact and as was shown in Ref. \cite{Badga17} this is fulfilled
for certain values of $Z$. These formal points are usually not analyzed
in the majority of works on the field. \\
ii)In addition to the resonances pole or direct contribution (normaly
referred as direct resonant or simply resonant, terms) to the amplitude,
we have background terms coming from cross resonance amplitudes (named
ussually background resonant terms) and non-resonance origin (called
usually background non resonant terms). Many works do not consider
the interference between these both resonant and background contributions
and really it is very important to describe the data. \\
iii)Finally, another models detach the decay process from the resonance
production out of the whole weak production amplitude. However, resonances
are nonperturbative phenomena associated to the pole of the S-matrix
amplitude and one cannot detach them from its production or decay
mechanisms. Further, the models to describe cross sections use born
contributions for the background amplitudes and resonant ones that
are valid around each resonance region. Nevertheless, the effect of
deviations from the hadronic pointlike couplings due to the quark
structure of nucleons and resonances are not taken into account when
the final $\pi N$ invariant mass $W_{\pi N}$ grows, and it is assumed
that that models are valid for all $W_{\pi N}$ regions. 

In this work we calculate the total and differential cross sections
of the inelastic dispersion of neutrinos on nucleons with the production
of one pion until $W_{\pi N}\lesssim2$ GeV and we try to describe
the reanalysed data in Refs.\cite{wilki14,Rodriguez16}. In our model
we incorporate explicitly resonance states $\Delta(1232)$, and $N^{*}(1440),N^{*}(1520),N^{*}(1535)$,
the so called second resonance region, with a consistent model for
the spin-$\frac{3}{2}$ fields that reproduced satisfactorily the
data for the ANL experiment in the range $W_{\pi N}\lesssim1.6$ GeV
\cite{Tamayo22}. The effect of hadron structure when $W_{\pi N}$
grows and of more energetic resonances, not introduced explicitly
in the model, is included throught a global effective hadronic FF
in consistence with previous calculations for neutral currents(NC)
\cite{Mariano11}. Our model reproduces well the data of Refs.\cite{wilki14,Rodriguez16},
both for the total and differential cross sections.

Our work will be organized as follows: In Section II, we summaryze
the general description of the pion production
cross sections, together with the previously calculated amplitudes showed in the Appendix.
In Section III we analyze how to extend our previous model to higher
$W_{\pi N}$ and will show our results for neutrino and antineutrino
scattering. Finally, in Section V we summaryze our conclusions.

\section{Total and differential cross sections}

We resume here some general concepts and let specific formulae for an Appendix. We are interested in describing here two observables for  the charged current (CC) $\nu N\rightarrow\mu^{+}N'\pi$ and $\bar{\nu}N\rightarrow\mu^{-}N'\pi$ modes, since neutral current (NC) processes in the present $W_{\pi N}$-range, where analyzed
previously \cite{Mariano11}.
The total cross section for single pion  weak production on nucleons, where we use  $\nu N$ center mass (CM) variables since it is easier to look for their limits, reads
(we take $\boldsymbol{p}_{\nu}=E_{\nu}\hat{\boldsymbol{k}}$ along
the z-axis)
\begin{eqnarray}
\sigma(E_{_{\nu}}^{\text{CM}}) & = & \frac{m_{_{\nu}}m_{_{N}}^{2}}{(2\pi)^{4}E_{_{\nu}}^{\text{CM}}\sqrt{s}}\int\limits _{E_{\mu}^{^{-}}}^{E_{\mu}^{^{+}}}dE_{_{\mu}}^{\text{CM}}\int\limits _{E_{\pi}^{^{-}}}^{E_{\pi}^{^{+}}}dE_{_{\pi}}^{\text{CM}}\int\limits _{-1}^{+1}d\cos\theta\int\limits _{0}^{2\pi}d\eta\frac{1}{16}\sum\limits _{\text{spin}}|\mathcal{M}|^{2},\label{eq:Totalcross}
\end{eqnarray}
being $\sqrt{s}=E_{_{\nu}}^{\text{CM}}+E_{_{N}}^{\text{CM}}=\sqrt{2E_{_{\nu}}m_{_{N}}+m_{_{N}}^{2}}$, $d\Omega_{\mu}=d\cos\theta d\phi$
and $d\Omega_{\pi}=dcos\xi d\eta$, and where $d\phi$ and $\cos\xi$  integrations are not present due to symmetry and  conservation fixing, respectively. The limits $E_{_{\mu}}^{\pm}$ and $E_{_{\pi}}^{\pm} $ can be found trivially from momentum conservation (see Ref.\cite{Barbero08}), and where the conexion with neutrino's energy in LAB is given by
\begin{eqnarray}
E_{_{\nu}}^{CM} & =&\frac{m_{_{N}}E_{_{\nu}}}{\sqrt{2E_{_{\nu}}m_{_{N}}+m_{_{N}}^{2}}},\label{eq:EcmLab}
\end{eqnarray}
being the LAB four momenta involved in Eqs.(\ref{eq:Totalcross}-\ref{eq:EcmLab}) defined as
\begin{eqnarray*}
p_{\nu}=(E_{\nu},{\bf p}_{\nu}),\hspace{0.5cm}p_{\mu}=(E_{\mu},{\bf p}_{\mu}),\hspace{0.5cm}\text{k}=(E_{\pi},{\bf k}),\hspace{0.5cm}p=(E_{N},{\bf p}),\qquad p'=(E_{_{N'}},{\bf p}'),
\end{eqnarray*}
with $E(\boldsymbol{\xi})=\sqrt{|\boldsymbol{\xi}|^{2}+m^{2}}$(we
set $m_{\nu}=0$), with the same expressions and definitions for $\bar{\text{\ensuremath{\nu}}}$.
On the other hand, since it is the tool for fixing axial resonance parameters by comparison with the ANL and BNL data experiments
\cite{Rad82,Kitagi86} of the neutrino flux  averaged
cross section, we will calculate the  flux averaged differential cross section defined as ($Q^{2}=-(p_{\mu}-p_{\nu})^{2}$)
\begin{eqnarray}
\left\langle \frac{d\sigma}{dQ^{2}}\right\rangle  & = & \frac{\int\limits _{E_{\nu}^{\text{min}}}^{E_{\nu}^{\text{max}}}\frac{d\sigma({E_{_{\nu}}})}{dQ^{2}}\phi(E_{\nu})dE_{\nu}}{\int\limits _{E_{\nu}^{\text{min}}}^{E_{\nu}^{\text{max}}}\phi(E_{\nu})dE_{\nu}},\label{eq:difcrossQ2}
\end{eqnarray}
being $\phi(E_{\nu})$ the neutrino's  flux.
The total amplitude $\mathcal{M}$ for the considered process reads

\begin{eqnarray}
\mathcal{M} & = &i\frac{G_{F}^{2}}{\sqrt{2}}\bar{u}(p_{\mu})(-)i\gamma^{\lambda}(1-\gamma_{_{5}})u(p_{\nu})ig_{\lambda\lambda'}V_{ud}\bar{u}(p')\mathcal{O}^{\lambda'}(p',k,p,q)u(p),\label{eq:amplitude}
\end{eqnarray}
being spin and isospin indexes omitted, $G_{F}=1.16637\times10^{-5}GeV{}^{-2}$
, $|V_{ud}|=0.9740$, and $ \mathcal{O}^{\lambda'}$ is the vertex generated by the hadronic CC current,  present in 
the  weak interaction Lagrangian  and the strong vertexes and hadron propagators (see Refs.
\cite{Barbero08,Mariano11}).
$\bar{u}O^{\lambda}u$ includes  the $WN\rightarrow\mu N'\pi$
processes present in Fig.1 and who defines the contribution of a  Feynman
graphs  to a channel, results to be  the isospin coefficients shown below. \medskip{}
\medskip{}
\medskip{}
\medskip{}
\medskip{}
\begin{figure}[h]
\includegraphics[scale=0.3]{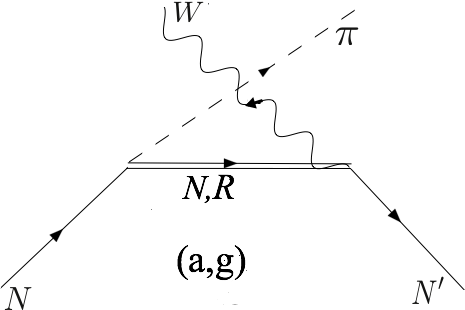}\includegraphics[scale=0.3]{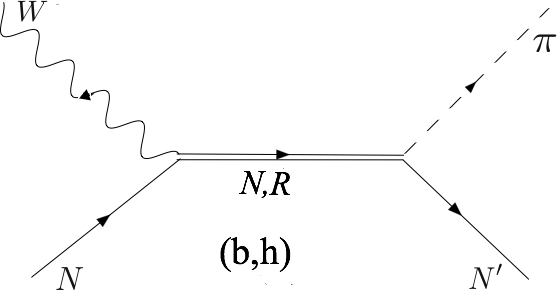}

\includegraphics[scale=0.3]{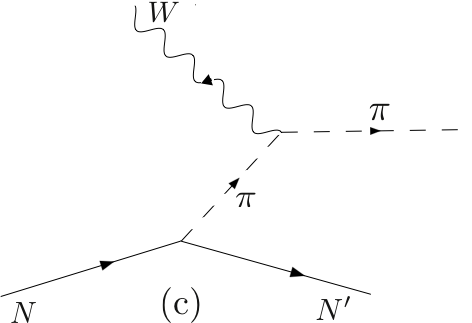}\includegraphics[scale=0.3]{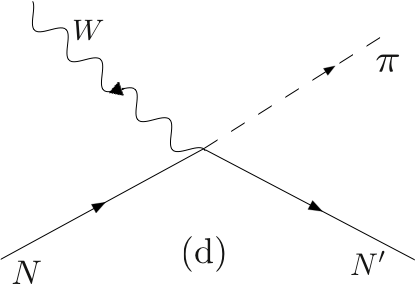}

\includegraphics[scale=0.3]{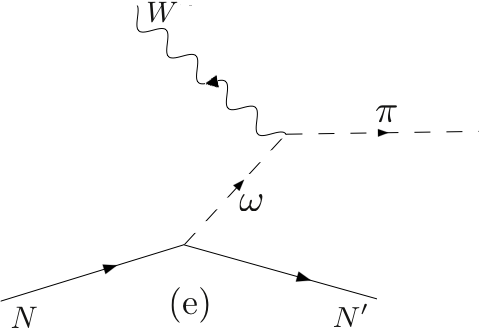}\includegraphics[scale=0.3]{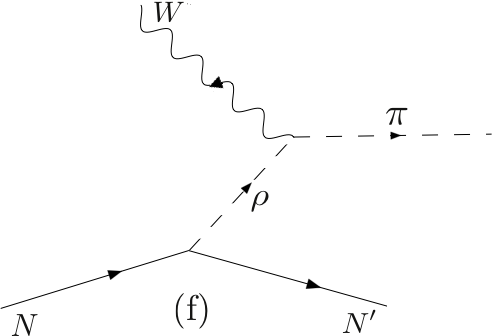}

\caption{Terms in  the scattering amplitude for the process $\nu(\bar{\text{\ensuremath{\nu}}})N\rightarrow\mu N'\pi$. Terms (a) and  (g)  or (b) and (h) respectively, have the same tophological form with a nucleon (N) or any resonance (R) as intermediate states. 
Fig (a)-(g) represent the background (B) contribution. Fig (h) is a the direct
or pole resonant contribution (R). }
\end{figure}
\medskip{}
\medskip{}

We can split $ \mathcal{O^{\lambda}}$ as
\begin{eqnarray}
\mathcal{O}^{\lambda} & = & \mathcal{O^{\lambda}}_{\text{B}}+\mathcal{O}_{\text{R}}^{\lambda},\label{eq:AmplitudB+R}
\end{eqnarray}
where $\mathcal{O}_{B,R}$ generate  the tree-level hadronic amplitudes  contributing to the background (B), nucleon Born terms (Fig. 1(a)-(b)),
the meson exchange amplitudes(contact term included) (Figs. 1(c)-(f))
and the resonance-crossed term (Fig. 1(g)); the direct or pole resonant contribution
(R) is shown in Fig. 1(h) and it is neccesary the resonance adquires
a width to  avoid the singulatity in the propagator by a self energy dressing that can be 
treated within  different  approximations. The effect of the self energy could, in principle, change the full structure of the propagator.
In the case of  spin-$\frac{1}{2}$
resonances the unstable character introduced by this  self energy accounts the replacement 

\begin{eqnarray}
m_{R} & \rightarrow & m_{R}-i\frac{\Gamma_{_{R}}(s)}{2},\label{eq:complexmass}
\end{eqnarray}
 into the unperturbed propagator without changing its structure. This width $\Gamma_R$ is obtained by considering the pion-nucleon loop contribution to the self energy \cite{Leitner09} and reads

\begin{eqnarray}
\Gamma_{R}(s)\times B_{r} & = & \frac{3}{4\pi}\left(\frac{f_{_{\pi NR}}}{m_{\pi}}\right)^{2}(m_{R}+\pi m_{N})^{2}\left(\frac{(\sqrt{s}-\pi m_{_{N}})^{2}-m_{_{\pi}}^{2}}{4s^{3/2}}\right)\lambda^{\frac{1}{2}}(s,m_{_{N}}^{2},m_{_{\pi}}^{2}),\label{eq:Gamma_s1/2}\\
\lambda(x,y,z) & = & x^{2}+y^{2}+z^{2}-2xy-2xz-2yz.\nonumber 
\end{eqnarray}
being $\pi=\pm$ the parity, $B_{r}$ the corresponding
$\pi N$ branching ratio decay,  and $f_{_{\pi NR}}$ the strong coupling. 

For   the  spin-$\frac{3}{2}$ resonances the inclusion of the pion-nucleon loop in the  self energy  alters
the full structure of the propagator due to the presence of spureous
$\frac{1}{2}$ components. Nevertheless, if we neglect  terms of order $\left(f_{_{\pi NR}}/{m_{\pi}}\right)^{4}$
and $((m_{R}-\sqrt{s})\left(f_{_{\pi NR}}/m_{\pi}\right)^{2})$ (see Ref.\cite{Barbero15}),
 expected to be very small  in the resonance
region $(\sqrt{s}\approx m_{R})$, we keep  the unperturbed form with the same replacement (\ref{eq:complexmass}) and now
\begin{eqnarray}
\Gamma_{_{R}}(s)\times B_{r} & = & \frac{\mathcal{I}_{R}\left(\frac{f_{_{\pi NR}}}{m_{\pi}}\right)^{2}}{4\pi}\left(\frac{(\sqrt{s}+\pi\times m_{_{N}})^{2}-m_{_{\pi}}^{2}}{48s^{5/2}}\right))\lambda^{\frac{3}{2}}(s,m_{_{N}}^{2},m_{_{\pi}}^{2}),\label{eq:Gamma_s}
\end{eqnarray}
where $\mathcal{I}_{R}=\left\{ \begin{array}{c}
1\\
3
\end{array}\right.$ for $I=\left\{ \begin{array}{c}
\frac{1}{2}\\
\frac{3}{2}
\end{array}\right.$ respectively. 

If one analyzes the formal scattering T-matrix theory
\cite{Mariano07} for the final $\pi N$ pair in both elastic scattering or pion photoproduction, it is mandatory that
the $\pi NR$ vertex should be also dressed as the propagator by the $\pi N$ rescattering through non-pole amplitudes.
This makes also the vertex  $s$-dependent, or in other words
we get an effective coupling constant $\frac{f_{_{\pi NR}}(s)}{m_{\pi}}$,
due to the decay \cite{Mariano07}
mediated by the intermediate $\pi N$-propagator (this will be analyzed
deeply below). In previous works it has been have considered the approach resulting from
the formal limit of massless $N$ and $\pi$ in the loop contribution
to the $\Delta$ self energy and in the dressed $\pi N\Delta$ vertex.
This was called the complex-mass scheme (CMS) \cite{Amiri92}. In this formal limit
the vertex dressing gives a dependence
$f_{_{\pi N\Delta}}(s)=\frac{\kappa f_{_{\pi N\Delta}}^{0}}{\sqrt{s}}$, being
$f_{_{\pi N\Delta}}^{0}$ the bare $\pi N\Delta$ coupling
constant and $\kappa$ a constant of dimension MeV to fit, in place of doing
the complex  calculation of the integral involved in the vertex correction.
Within the CMS we derive from (\ref{eq:Gamma_s}) the approximated
expression ($B_{r}\approx1$ for the $\Delta$)
\begin{eqnarray}
\Gamma_{\Delta}(s) & = & \left(1-\frac{\sqrt{s}-m_{R}}{m_{R}}\right)\Gamma_{\Delta}^{CMS},\;\Gamma_{\Delta}^{CMS}=\frac{\kappa^{2}(\frac{f_{_{\pi NR}}^{0}}{m_{\pi}})^{2}}{192\pi}m_{\Delta}.\label{eq:CMS}
\end{eqnarray}
When  $s\simeq m_{\Delta}^{2}$ we get a constant width $\Gamma_{\Delta}(s)\approx\Gamma_{\Delta}^{CMS}$,
where  $\Gamma_{\Delta}^{CMS}$ is fitted in place of $\kappa$
together $\frac{f_{_{\pi N\Delta}}}{m_{\pi}}$ and $m_{\Delta}$ to
reproduce $\pi^{+}p$ scattering \cite{Mariano01}. Another approach
commonly used\cite{Leitner09}, is to fix $\sqrt{s}\approx m_{\Delta}$
in (\ref{eq:Gamma_s1/2}) or (\ref{eq:Gamma_s}) and to use the experimental
values for $m_{\Delta}$ and $\Gamma_{\Delta}$ times $B_{r}$, and
get $f_{_{\pi NR}}$. We will refer to this as constant mass-width
approach (CMW). We will use both the CMS and CMW depending on the
considered resonance. 

The  explicit expressions for $\mathcal{O}_{B}$, splitted in those coming from the nucleon and meson exchange contributions (BN) and from the resonances(BR), and  $\mathcal{O}_{R}$ for  $R\equiv \Delta(1232)$, $N^{*}(1440),N^{*}(1520),N^{*}(1535)$ are shown in the Appendix. These are obtained in Ref.\cite{Tamayo22}. We only shown here the isospin factors for each contribution to the amplitude (see Fig.(1)) since it will be useful for the results section.  They  are  indicated with
$\mathcal{T}(m_{t}\equiv p,n,n,m_{t'}\equiv p,p,n )$(or charge conjugate for $\bar{\nu })$ and can be obtained using the isospin operators present in the interaction Lagrangians togheter the isospin wave funcions for the $W$ boson and the hadrons (see Refs. \cite{Mariano11} and \cite{Barbero08}  )

\begin{align}
\mathcal{T}_{a} & =\mathcal{T}_{g}^{1440,1535,1520}=-2,0,-\sqrt{2}\nonumber \\
\mathcal{T}_{b} & =\mathcal{T}_{h}^{1440,1535,1520}=0,-2,\sqrt{2}\nonumber \\
\mathcal{T}_{c} & =1,-1,\sqrt{2}\nonumber \\
\mathcal{T}_{d}& =-1,1,-\sqrt{2}\nonumber \\
\mathcal{T}_{e}& =-1,-1,0\nonumber \\
\mathcal{T}_{f}& =1,1,-\sqrt{2}\nonumber \\
\mathcal{T}_{g}^{\Delta}& =-1/3,-1,\sqrt{2}/3\nonumber \\
\mathcal{T}_{h}^{\Delta}& =-1,-1/3,-\sqrt{2}/3.\label{eq:isospincoef}
\end{align}

\section{Extension to higer $W_{\pi N}$ and and Results}

In this work we analyse the total and differential cross sections
for the charged current (CC) modes of the six processes
\begin{eqnarray}
\nu p & \rightarrow & \mu^{-}p\pi^{+},\hspace{0.4cm}\nu n\rightarrow\mu^{-}p\pi^{0},\hspace{0.4cm}\nu n\rightarrow\mu^{-}n\pi^{+},\nonumber \\
\bar{\nu}n & \rightarrow & \mu^{+}n\pi^{-},\hspace{0.4cm}\bar{\nu}p\rightarrow\mu^{+}n\pi^{0},\hspace{0.4cm}\bar{\nu}p\rightarrow\mu^{+}p\pi^{-},\label{eq:process}
\end{eqnarray}
with neutrino energies exciting the $W_{\pi N}=\sqrt{(p'+k)^{2}}\leq2$ GeV
region. We will obtain these cross sections through the Eqs.(\ref{eq:Totalcross}-\ref{eq:difcrossQ2})
with the amplitude (\ref{eq:amplitude}), taking $(1+\gamma_{5})$
when $\text{\ensuremath{\nu}}\rightarrow\bar{\nu}$, and the vertex
production contributions in Eqs.(\ref{eq:OBN}-\ref{eq:OR}) of the Appendix. We have
included explicitly the resonances of the first and second region
($\Delta(1232),$$N^{*}(1440),N^{*}(1520),N^{*}(1535)$), while the
effect of more energetic ones will be included indirectly. A calculation
with the cuttoff $W_{\pi N}\lesssim1.6$ GeV using the above mentioned
CMS approach for the $\Delta(1232)$ and CMW for the other resonances,
described properly $\sigma,\frac{d\sigma}{dQ^{2}},\frac{d\sigma}{dW_{\pi N}}$
\cite{Tamayo22} for the ANL data. We have treated in consistent fashion
the spin-$\frac{3}{2}$ resonances from the point of view of contact
invariance and the true considered degree of freedom of the spin-$\frac{3}{2}$
field, both points mentioned in the introduction, choosing the values
$A=-\frac{1}{3},Z=\frac{1}{2}.$ Now, we wish to extend our model
to higher energies in order to describe the data of the recent reanalyzed
ANL and BNL data without cuts\cite{Rodriguez16}. On going to higher
energies we can ask ourselves two important questions. First, is it
possible to pursue the tree level model used for the background or
the CMS and CMW approches to treat the resonances for any energy $W_{\pi N}$
with pointlike hadrons? Second, it is enough simply to add more and
more resonances to describe the $W_{\pi N}>1.6$ GeV region? We will
intend to analyze these questions within the frame of $\pi N$ rescattering,
not considered in detail until now, in the following subsections.

\subsection{Rescattering and hadrons FF}

We support our discusion on the pion photoproduction reaction analyzed
in Ref.\cite{Mariano07}, and we have the analogy $\mathcal{O}^{\lambda}\equiv\hat{B}^{\lambda}$
and $\mathcal{O}_{B,R}^{\lambda}\equiv\hat{B}_{NP}^{\lambda},\hat{B}_{P}^{\lambda}$
with the photoproduction vertexes in that reference. The approximation
implemented previoulsy until the second resonance region\cite{Tamayo22}
can be resummed as 

\begin{eqnarray}
\mathcal{O}^{\lambda} & \approx & \mathcal{O}_{B}^{\lambda}+\mathcal{O}_{R}^{\lambda}\equiv\mathcal{O}_{B}^{\lambda}+\sum_{R}V_{R\pi N}G_{R}W_{WNR}^{\lambda},\label{eq:aproxamp}
\end{eqnarray}
where $V_{R\pi N}$ and $W_{WNR}^{\lambda}$ represent schematically
the $R\rightarrow N\pi$ and $WN\rightarrow R$ vertexes respectively,
while $G_{R}$ a resonance propagator. Nevertheless, the $\mathcal{O}_{B}^{^{\lambda}}$
contributions in special the $\mathcal{O}_{BR}^{\lambda}$ coming
from the $\frac{3}{2}$ resonances (Fig.1(g)) can not be dressed by
the self energy, being not affected by a $\pi N$ rescattering in
(\ref{eq:aproxamp}) as done in $\mathcal{O}_{R}^{\lambda}$ with
$\Gamma_{R}$. As consequence, grows rapidly for $W_{\pi N}\gtrsim1.5$ GeV
as will be shown below. As described in Ref.\cite{Mariano07} for
the case of photoproduction, but valid also here, there are other
effects not considerated. The complete excitation amplitude in the
final $\pi N$ CM is 

\begin{eqnarray}
\mathcal{O}^{\lambda}(k,q) & = & \left[-i(2\pi)^{4}\delta^{4}(k-k')+T_{NP}(k')G_{\pi N}(k')\right]\left(\mathcal{O}_{B}^{\lambda}(k',q)+\sum_{R}V_{R\pi N}(k',k)G_{R}(k)\widetilde{W}_{WNR}^{\lambda}(k,q)\right)\label{eq:exactamp}\\
\widetilde{W}_{WNR}^{\lambda}(k,q) & = & W_{WNR}^{\lambda}(k,q)+\left(V_{R\pi N}(k,k'')+V_{R\pi N}(k,k')G_{\pi N}(k')T_{NP}(k',k'')\right)G_{\pi N}(k'')\mathcal{O}_{B}^{\lambda}(k'',q)\label{eq:Rvertex}
\end{eqnarray}
where $k',k''$ are intermediate pion momenta and a repeated $k,k',k''$
are indicate $i\frac{\int dk,k',k''^{4}}{\left(2\pi\right)^{4}}$
, 

\begin{eqnarray*}
G_{\pi N}(k') & = & S_{N}(p+q-k')\Delta_{\pi}(k'),
\end{eqnarray*}
is the pion-nucleon intermediate propagator, and $T_{NP}$ the non-pole
scattering T-matrix that iterates to all orders the potential $V_{NP}\equiv V_{N\pi,N'\pi'}$
built with nucleon Born terms, meson exchange t-contributions and
u-resonant contributions. In summary, in the full amplitude the rescattering
of the final $\pi N$ pair through $T_{NP}$ is considered as well
the decay into a resonance of $\mathcal{O}_{B}^{\lambda}$. We will
not introduce in this work unitarizartion corrections, done through
imaginary contributions in (\ref{eq:exactamp}), since as we are analyzing
the total and differential cross sections, where corrections to each
multipole compensate out in the multipole expansion of the cross section\cite{Mariano07}.
As we wish to deal with effective real coupling constants ($T_{NP}$
dresses but is a complex operator), it is convenient to express the
$\pi N$ T-matrix operator in terms of the real $K$-matrix \cite{Mariano07},
and after a three dimensional reduction we get\cite{Mariano07}

\begin{eqnarray}
\mathcal{O}^{\lambda}(\boldsymbol{k},\boldsymbol{q},W_{\pi N}) & \approx & \left[(2\pi)^{3}\delta^{3}(\boldsymbol{k}-\boldsymbol{k'})+\mathcal{P}\left(K_{NP}(\boldsymbol{k},\boldsymbol{k}')G_{TH}(\boldsymbol{k}',\sqrt{s})\right.\right]\nonumber \\
 & \times & \left[\left.\mathcal{O}_{B}^{\lambda}(\boldsymbol{k}',\boldsymbol{q},W_{\pi N})\right)+\sum_{R}\left.V_{R\pi N}(\boldsymbol{k}',\boldsymbol{p})\right)G_{R}(\boldsymbol{p},W_{\pi N})\widetilde{W}_{WNR}^{\lambda}(\boldsymbol{p},\boldsymbol{q},W_{\pi N})\right],\label{eq:PVamplitude}\\
\widetilde{W}_{WNR}^{\lambda}(\boldsymbol{p},\boldsymbol{q},W_{\pi N}) & = & W_{WNR}^{\lambda}(\boldsymbol{p},\boldsymbol{q})+\mathcal{P}\left[V_{R\pi N}(\boldsymbol{p},\boldsymbol{k'})G_{TH}(\boldsymbol{k'},W_{\pi N})\mathcal{O}_{B}^{\lambda}(\boldsymbol{k'},\boldsymbol{q})\right],\label{eq:WPV}
\end{eqnarray}
where $G_{\pi N}(k')$ is replaced by the Thompson propagator $G_{TH}(\boldsymbol{k'},W_{\pi N})=\frac{m_{N}}{2E_{\pi}(\boldsymbol{k'})E_{N}(\boldsymbol{k'})}\sum_{ms'}\frac{u(-\boldsymbol{k'},m_{s}')\bar{u}(\boldsymbol{k'},m_{s}')}{W_{\pi N}-E_{\pi}(\boldsymbol{k'})-E_{N}(\boldsymbol{k'})}$
and where with $\mathcal{P}$ is the principal value on the integral
in reapeated momenta. 

In order to reproduce the experimental data FF have to be introduced
which affect both $\mathcal{O}_{B,R}^{^{\lambda}}$ to regularyze
the integrals in (\ref{eq:PVamplitude}) and (\ref{eq:WPV}) . They
are meant to model the deviations from the pointlike couplings due
to the quark structure of nucleons and resonances, analogs of the
electromagnetic ones reflecting the extension of the hadrons, and
should be calculated from the underlying theory or quark models\cite{Melde09}.
Because it is not clear a priori which form these additional factors
should have, they introduce a source of systematical error in all
models\cite{Mosel98}. Guided by an our previous proper description
on NC1$\pi$ data obtained by the CERN Gargamelle experiment without
applying cuts in the neutrino energies\cite{Mariano11}, we multiply
$\mathcal{O}_{B}^{\lambda}+\mathcal{O}_{R}^{\lambda}$ by a global
regularyzing FF of the $R\pi N$ and $N\pi N'$ vertexes (${\rm k}=|\boldsymbol{k}|$)

\begin{eqnarray}
F({\rm k},W_{\pi N}) & = & \frac{\left(\Lambda\right)^{4}}{\left(\Lambda\right)^{4}+{\rm k(W_{\pi N})^{2}}\left(W_{\pi N}-W_{\pi N}^{th}\right)^{2}\theta(W_{\pi N}-1.6\mbox{{GeV}})},\label{eq:FF4mod-1}\\
\mathrm{\:k}(W_{\pi N}) & = & \sqrt{\frac{(W_{\pi N}^{2}-m_{N}^{2}-m_{\pi}^{2})^{2}-4m_{N}^{2}m_{\pi}^{2}}{4W_{\pi N}^{2}}},
\end{eqnarray}
being the threshold invariant mass $W_{\pi N}^{th}=m_{\pi}+m_{N}$
and which is consistent with that introduced in Refs.\cite{Sato96}
and \cite{Pearce91}(\cite{footnote}), but lighting it above the second resonance region since it was shown
that the description with the $W_{\pi N}\lesssim1.6$ GeV was correct
within our model \cite{Tamayo22}. This FF can be seen as 

\begin{eqnarray}
F({\rm k},W_{\pi N}) & = & \frac{\left(\Lambda_{eff}\right)^{2}}{\left(\Lambda_{eff}\right)^{2}+{\rm k(W_{\pi N})^{2}}},\Lambda_{eff}=\Lambda\frac{\Lambda}{W_{\pi N}-W_{\pi N}^{th}},\label{eq:effmonopole}
\end{eqnarray}
where we have a monopole FF with an effective cutoff diminishing with
$W_{\pi N}-W_{\pi N}^{th}$, making that certain term ''disappears''
or contributes less in the amplitude since when $W_{\pi N}$ grows
another resonance, not considered in the amplitude, could be excited.
Note that this FF affects also the on-shell contributions,i.e, the
terms surviving in (\ref{eq:PVamplitude}) when the $\mathcal{P}$
terms are dropped.

In resume, to get the full amplitudes in Eq.(\ref{eq:PVamplitude})
we need to add FF taking into account the hadron extensions since
as can be seen in $G_{TH}(\boldsymbol{k'},W_{\pi N})$, we keep the
$\pi$ and $N$ elemental character for any $W_{\pi N}$, and this
makes the involved integral divergent. This entails to moderate the
on shell amplitude $\mathcal{O}^{\lambda}$ with this FF, which is
not considered in the approach (\ref{eq:aproxamp}).

\subsection{Results }

We adopt the same model as in a recent work\cite{Tamayo22}, where
we have calculated the pion production cross section including explicitly
spin-$\frac{1}{2}$ and $\frac{3}{2}$ resonances $\Delta(1232),N^{*}(1440),N^{*}(1520)$
and $N^{*}(1535)$ in the calculus of the amplitudes(\ref{eq:OBN}),(\ref{OBR})
and (\ref{eq:OR}) in the Appendix to cover the so called second resonance region.
The formal aspects mentioned in the introduction an related with the
spin-$\frac{3}{2}$ Lagrangians (free and interaction) regards the
contact-transformation parameter and the additional $Z$ parameter
present in the $\mathcal{L}_{\pi NR}$ interaction lagrangian, have
been discussed in that reference. The non resonant contributions in
(\ref{eq:OBN}) are also described in \cite{Tamayo22} together all
the adopted FF. We treat the spin-$\frac{1}{2}$ resonances within
the parity conserving parametrization for the FF, since this is compatible
to that used in the similar topologycal nucleon contribution in Fig.1
(a) and (b). For the spin-$\frac{3}{2}$ resonances we use the Sachs
parametrization to be consistent with our previous works including
only the $\Delta$ resonance, where we get better results than using
the parity conserving one\cite{Barbero14}. We followed the conexion
between both parametrizations achieved in that reference to get the
FF for the $N^{*}(1520)$ resonance, and have taken the $Q^{2}$-depedent
FF from Ref.\cite{Lakakulich06} for all the second region resonances.
The main channel $\nu p\longrightarrow\mu^{-}\pi^{+}p'$ with the
cut $W_{\pi N'}<1.4$ GeV in the final invariant mass, is used to
fix the main axial $\Delta$-coupling constant $D_{1}(0)$ Refs.\cite{Mariano07,Barbero08}.
With this cut it was shown \cite{Tamayo22}, at less for this channel,
that the contributions of more energetic resonances than the $\Delta(1232)$
are small and that are important only for more energetic cuts. As
the reanalyzed data of ANL achieved in Ref.\cite{Rodriguez16} does
not affect appreciably the channel used to fit $D_{1}(0)$ for $W_{\pi N'}<1.4$
GeV, then we will not make a new fitting to $\left\langle \frac{d\sigma}{dQ^{2}}\right\rangle $.
In our previous work \cite{Tamayo22} we achieved the comparison with
the data of ANL experiment in the region of $W_{\pi N}<1.4,1.6$ GeV,
where we have worked within the CMS+CMW approach. From the results
including and not including the second resonance region, we concluded
that to describe the data in all channels this resonance region should
be included. We concluded that these resonances influence through
their tails, since they have their centroids over these cuts. At the
amplitude level, these tails interfere with the another resonances
and background contributions. This behavior is confirmed when we compared
with the data with the $W_{\pi N}<1.6$ GeV and the good agreement
with the data for the antineutrino case.

\begin{figure}[h]
\textsf{\large{}\includegraphics[width=20cm,height=13cm]{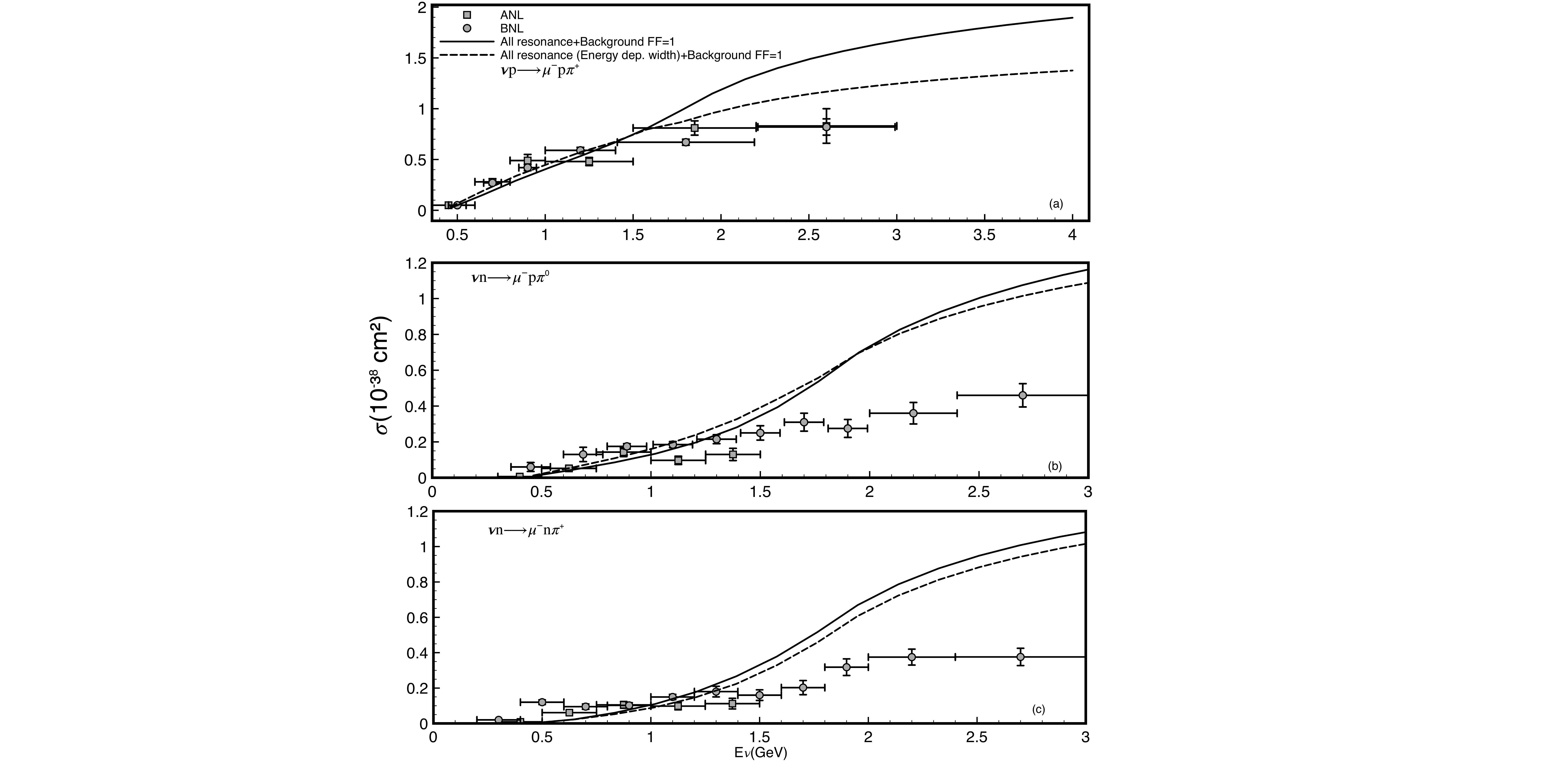} \caption{Total $\nu N$ cross section as function of the neutrino energy for
different channels and $F({\rm k},W_{\pi N})=1$. Results with $\Delta+$second
region resonances plus the corresponding background in each case are
shown for a cut $W_{\pi N}\lesssim2$ GeV with dotted and full lines.
Results within CMS+CMW and with energy dependent width with are shown
with dashed lines. Data are taken form Ref.\cite{Rad82}}
\label{FF=00003D1} }
\end{figure}

Nevertheless, the ANL data of Ref.\cite{Rad82} contains also results
without energy cuts and also all results in Ref.\cite{Kitagi86} are
reported without events exclusion. In addition, the reanalysis of
these two set of data has been done recently in Refs.\cite{wilki14,Rodriguez16}
where the main results of the cross section are obtained with a tunable
invariant mass value, which is W \ensuremath{\le} 1.7 GeV by and compared
with data without cuts. For describing them we need to extend our
model to higher energies. The main question seems to be, do we add
more energetic resonances keeping the approach in Eq.(\ref{eq:aproxamp})?
or do we put attention on the FF discussion in the previous subsection?
In the Fig.(\ref{FF=00003D1}) we show the total cross section (\ref{eq:Totalcross})
calculated with the amplitude (\ref{eq:amplitude}) obtained from
the approach (\ref{eq:aproxamp}) and Eqs.(\ref{eq:OBN}-\ref{eq:OR}) in the Appendix.
Now we enable that $W_{\pi N}\lesssim2$ GeV, as can be seen the cross
section grows with a departure from the data. With full lines we show
the results of the CMS + CMW approach for the resonances with a constant
width, while with dashed lines we show results used the same masses
$m_{R}$ and coupling constants $f_{\pi NR}$ but with energy dependent
widths given by Eqs.(\ref{eq:Gamma_s1/2}) and (\ref{eq:Gamma_s}).
As can be seen a better width's approach, what means a most exact
treatment in the resonance self energy, improves appreciably the description
for the first channel. In it, the direct contribution (Fig.1(h)) for
the $\Delta$ with$\mathcal{T}_{h}^{\Delta2}=1$ in Eq.(\ref{eq:isospincoef}),
is the main contribution being the cross one with $\mathcal{T}_{g}^{\Delta2}=1/9$
very small. For the other two channels, this contribution is much
lower since $\mathcal{T}_{h}^{\Delta2}=1/9,2/9$ and the background
contribution with $\mathcal{T}_{g}^{\Delta2}=1,2/9$ is the most or
at least equal important.

\textsf{\large{}}
{\large \par}

These cross terms contribute to the background and cannot dressed
by the $\Delta$ self energy. The same kind of analysis can be done
for the other $\frac{1}{2}$-resonances, that have the same isospin
factors that the direct and cross nucleon terms (Figs.1(b) and (a)),
but with a much lower contribution of these regards the $\Delta$.
This analysis is an indication that to include rescattering effects
could be more important that to add more and more energetic resonances,
since background contributions will grow anyway. In resume, all background
contributions coming from non-resonant origin plus cross-resonant
terms (Figs.1(a)-1(g)) cannot be dressed by a self energy and its
behavior could be corrected taking into account the rescattering through
$T_{NP}$ as discussed in the previous subsection. 

As was mentioned, in Eqs. (\ref{eq:PVamplitude}) and (\ref{eq:WPV})
$\mathcal{O}^{\lambda}$ should be effected by a FF.
As to solve numerically the $\mathcal{P}$-integrals is not trivial,
we will assume the minimum approach of considering :\medskip{}

a) The use of effective couplings in $W_{WNR}^{\lambda}$, that is
$\widetilde{W}_{WNR}^{\lambda}\rightarrow W_{WNR}^{\lambda}(\mbox{{effective\:weak\:couplings}})$
to simulate the effect of the second term of Eq.(\ref{eq:WPV}), that
was shown satisfactory for the case of photoproduction\cite{Mariano07}.
These effective values are the empirical one adopted in our previous
work \cite{Tamayo22}. Same will be valid for $\mathcal{O}_{B}^{\lambda}$
where weak empirical or experimental coupling constants are used.\medskip{}

b)To avoid model dependencies coming from the introduction of arbitrary
FF at each interaction vertex, we will introduce a global form factor
(\ref{eq:FF4mod-1}) in the decay amplitude. The using of FF at each
vertex requires the introduction of vertex corrections to keep for
example electromagnetic gauge invariance. As we are including resonances
with effects ultil around $W_{\pi N}\sim1.6$ GeV, taking into account
the width of the most energetic considered one($S_{11}(1535)$ MeV),
we will only light on this FF above this energy. 

In this way, we also correct the amplitude for excitation of more
energetic resonances not considered, through the effective monopole
form in Eq.(\ref{eq:effmonopole})(see the followed discussion), in
our model since now we wish describe data without $W_{\pi N}$ cuts.
Guided by a previous proper description on NC1$\pi$ data obtained
by the CERN Gargamelle experiment without applying cuts in the neutrino
energies\cite{Mariano11}, we multiply $\mathcal{O}_{B}^{\lambda}+\mathcal{O}_{R}^{\lambda}$
by a global regularyzing FF of the $\Delta\pi N$ and $N\pi N'$ vertexes
described above in Eq.(\ref{eq:FF4mod-1}). We adopt the value $\Lambda=600$
MeV in consistence with Ref.\cite{Mariano11}. Also we throw out the
$\mathcal{P}$ contribution in the first bracket of Eq.(\ref{eq:PVamplitude})
modifying only the onshell potential, as a first minimum modification
to the model and that was appropiated when we discussed NC1$\pi$\cite{Mariano11}.
In resume, this FF should take into account the hadron structure and
the posibility of exciting more energetic resonances, when $W_{\pi N}$
grows, in an effective fashion. 

\textsf{\large{}\negthickspace{}\negthickspace{}\negthickspace{}\negthickspace{}\negthickspace{}\negthickspace{}\negthickspace{}\negthickspace{}\negthickspace{}\negthickspace{}\negthickspace{}\negthickspace{}\negthickspace{}\negthickspace{}\negthickspace{} }{\large \par}

\textsf{\large{}}
\begin{figure}[h!]
\textsf{\large{}\vspace{-1cm}
}{\large \par}

\textsf{\large{}\includegraphics[width=15cm,height=13cm]{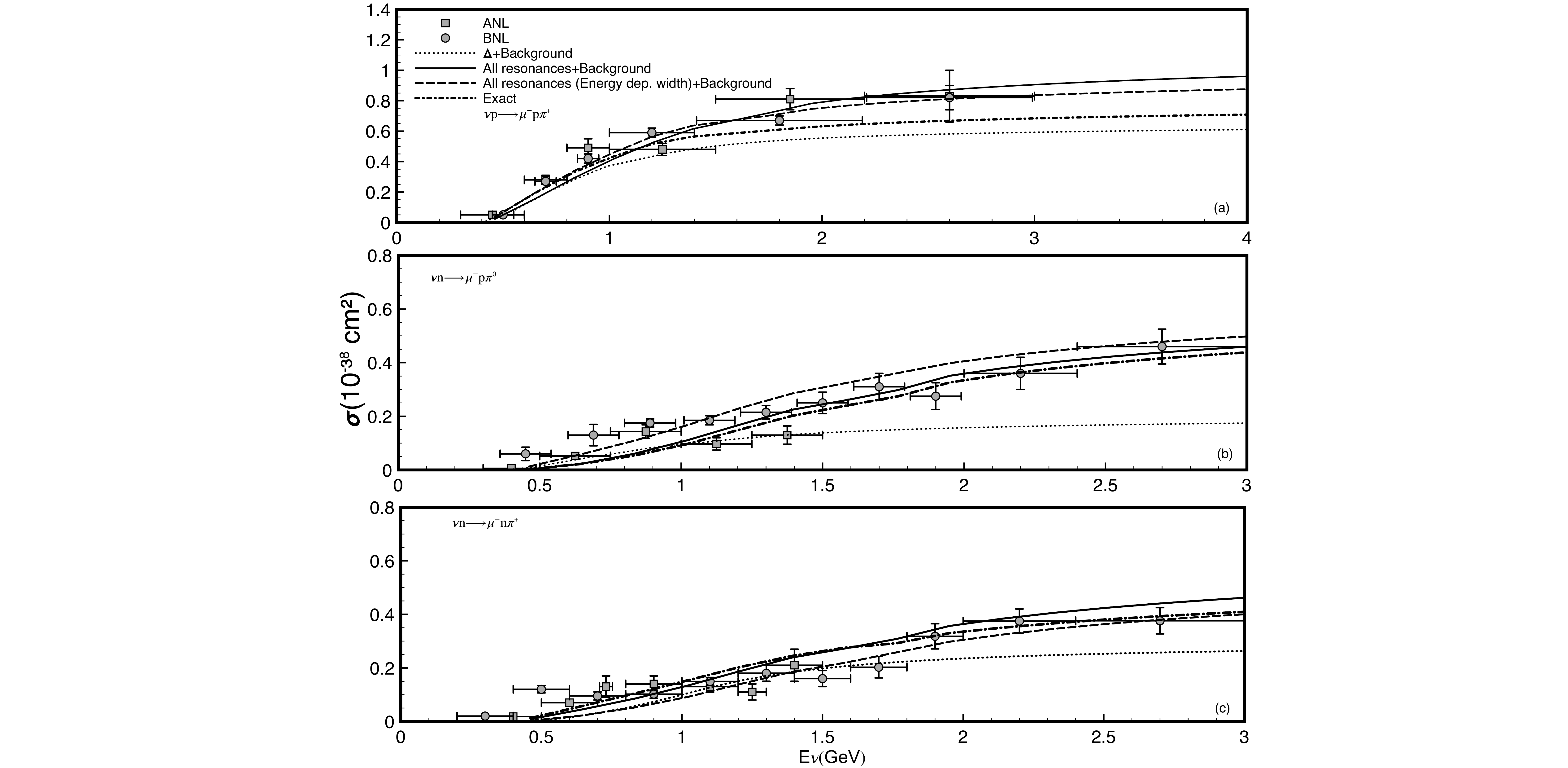}
\caption{Total $\nu N$ cross section as function of the neutrino energy for
different channels. Results with only the $\Delta$ and all resonances
= $\Delta+$second region's resonances, plus the corresponding background
are shown for $W_{\pi N}\lesssim2$ GeV. Data are taken from the reanalyzed
ANL and BNL ones in Ref.\cite{Rodriguez16}. $\Delta+$background
in CMS, dotted lines. All resonances + background in CMS+CMW, full
lines. Same but with energy dependent width from Eqs.(\ref{eq:Gamma_s1/2},\ref{eq:Gamma_s}),
dashed lines. Finally, same but using the exact expresion for the
$\Delta$ propagator (see. Ref.\cite{Barbero14}), dashed-dotted lines}
\label{nocut} }{\large \par}
\end{figure}
{\large \par}
Now, we discuss with more detail the calculations with $W_{\pi N}\lesssim2$
GeV and compare with data without cuts. As we have seen in the previous
Fig.(\ref{FF=00003D1}) the model used to treat the self energy in
the propagators leads to different results. Firstly, we will use the
CMS+CMW approaches that assume a constant width, secondly kepping
the same propagators but with the energy dependet width in Eqs.(\ref{eq:Gamma_s1/2})
and (\ref{eq:Gamma_s}), and finally we assume the exact $\Delta$(that
is the main contributing one) propagator. In this last case, the self
energy change the structure of the $\Delta$ propagator\cite{Barbero14}
but we assume the simplification of using  the same effective mass
and width of the the CMS approach. Results are shown in the Figure(\ref{nocut}).
As can be seen, the tendency of increasing the cross section by the
second resonance region contribution is persistent as previously \cite{Tamayo22},
and the results that better reproduce globally the three channels
the data is still the CMS+CMW approach with constant width. When we
enable an energy dependent width, we see a change that depends on
the channel diminishing the results in the main one (first pannel
of Fig.(\ref{nocut} )) and growing and diminishing in the other two
(second an third pannels of Fig.(\ref{nocut} )). When the exact propagator
for the $\Delta$ is used, the results for the first channel is more
diminished and an opposite effect is generated for the other two one.
\vspace{-0.25cm}
\singlespacing
Nevertheless, an observation should be done in order. The parameters used for the $\Delta$ resonance 
($f_{\pi N\Delta},m_{\Delta},\Gamma_{\Delta},$ $G_{E},G_{M},D_{1}\mbox{ or }C_{A}^{5}$) have been fitted using the CMS approach with a constant width, 
and thus if one wish to use another approach a new fitting should be done.This could explain why the best fit is done with the simpler CMS approach,
while for the energy dependent width should be not crucial a change
of parameters since we are keeping the same structure of the CMS propagator.
In addition, as we are using a global FF that affects the full amplitude,
and if the other resonances are treated within the CMW approach, to
use an exact propagator for the $\Delta$ would be no so consistent.
Finally, it is evident that the FF taken from NC pion productions
also work\emph{s} very well here.\textsf{\large{}} 
\singlespacing
\begin{figure}[h!]
\textsf{\large{}\negthickspace{}\negthickspace{}\negthickspace{}\negthickspace{}\negthickspace{}\negthickspace{}\negthickspace{}\negthickspace{}\negthickspace{}\negthickspace{}\includegraphics[width=20cm,height=16cm]{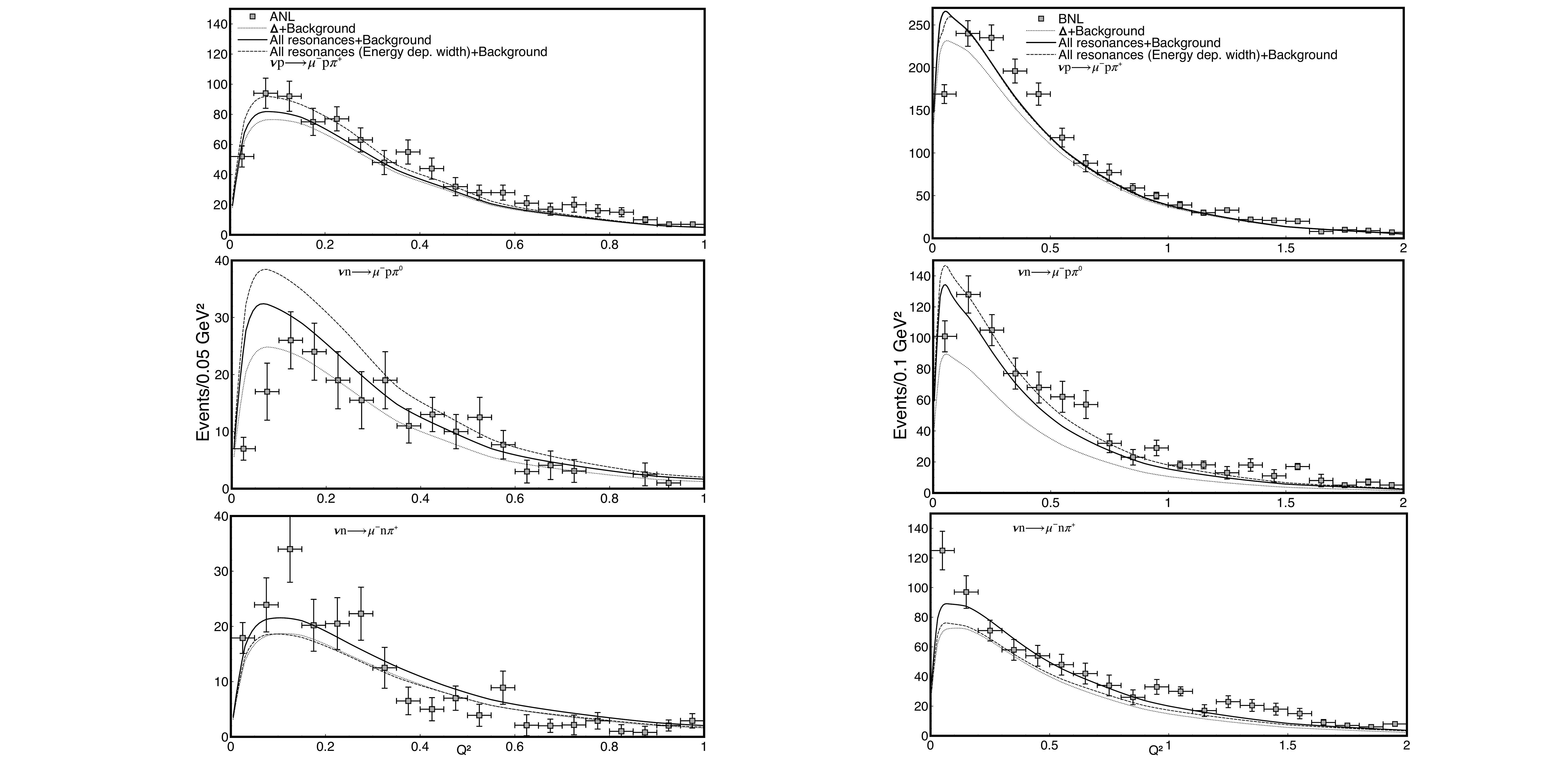}
\caption{Differential cross section with $W_{\pi N}\lesssim2GeV$. Lines convention
are the same that in previous Fig.(\ref{nocut}). Data are taken from
Ref.\cite{Rodriguez16}. }
\label{dQ2BNL} }
\end{figure}
{\large \par}

 \vspace{-0.35cm}This is  supported by  the following analysis. We have shown previously in the Ref. \cite{Mariano11} , from where the parameter $\Lambda$  in Eq.\rf{eq:effmonopole} was taken, the effect of the uncertainties in the parameters $D1(0)$ and $M_{A}$ in the axial FF  for the $\Delta$ resonance within the CMS model through a hatched area in the Figs.(7) and (8). There, it was shown that the uncertainties effects on the results is well below the data errors and the differences when we change from energy dependent  to constant widths approaches. These two parameters where involved in the fitting we done on the 
flux averaged $d\sigma/dQ^{2}$ cross section in the previous Ref. \cite{Barbero08} and then used in the following contributions, while the axial coupling for the another resonances were fixed from the PDG values for masses and widths. As we are not doing any new fitting here but using the parameters of previous works, we felt not necessary to analyzed the CMS+CMW parameters uncertainties effects again, because they are under control.

As the fitting of the axial $\Delta$ parameters it is achieved using
the flux averaged diferential cross section $<d\sigma/dQ^{2}>$, we
show the results for it within our model and compare with the more
recent ANL and BNL reanalyzed data \cite{Rodriguez16}. In that reference
results are shown for the events-$Q^{2}$ distribution in units of
Events/GeV$^{2}$ and as our results are given in $10^{-38}$cm$^{2}/$GeV$^{2}$
we use the same conversion factor found in the total cross section
calculation to compare with the data. They are reported without cuts
and are compared with our results in Fig.(\ref{dQ2BNL}). We show
calculations within the same approaches than the total cross section,
but avoid to use the exact $\Delta$ propagator for the reasons exposed
above. As can be seen, results within the constant width are acceptable
in the three channels. We are not doing any new fit (the fixing of
$C_{A}^{5}$was done previuolsy \cite{Mariano01} using the ANL data
with cuts $W_{\pi N}<1.4$ GeV\cite{Rad82}), and the description
is  better that those done with GENIE in Ref. \cite{Rodriguez16}.
As can be seen the using of energy dependent width enlarges the first
and second channels theoretical results and diminishes those for the
third one, this leads to a worse coincidence with data in an amount
depending on the experiment ANL or BNL.

\subsection{ANTINEUTRINOS}

\begin{figure}[h!]
\includegraphics[width=15cm,height=13cm]{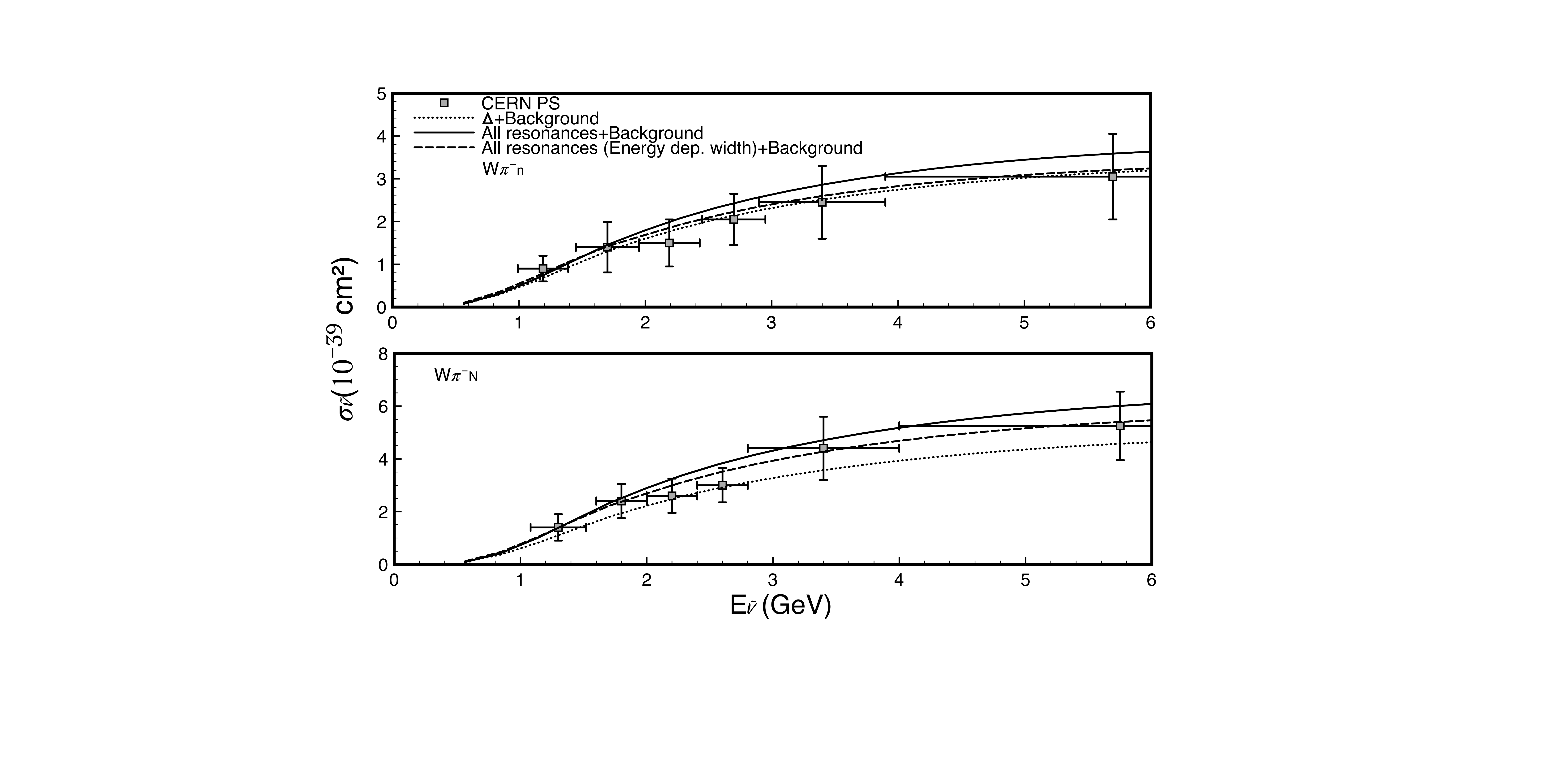}
\vspace{-2cm}
\caption{\label{fig:Antinunocut}Antineutrino's total cross sections with a
$W_{\pi N}\lesssim2$ GeV for the $\bar{\nu}n\rightarrow\mu^{+}n\pi^{-}$
and that leadding to a final $N\pi^{-}$final state. Lines conventions
are the same that in Fig.(\ref{nocut}).}
\end{figure}

Now, in order to follow probing our model we wish to calculate the
antineutrinos total cross section as done in our previous work\cite{Tamayo22}.
We have to differences regards the neutrinos case. Firstly the interactions
of neutrinos with hadrons is not the same that for antinetrinos. We
have a sign of difference in the lepton current contraction that makes
a different coupling with the hadron one. Then, the interaction with
neutrinos is different from antineutrinos due the use of spinors for
antiparticles in the lepton current in Eq.(\ref{eq:amplitude}) and
has nothing to do with the very know CP violation. Secondly, in the
experiment an admixture of heavy freon CF3Br was exposed to the CERN
PS antineutrino beam (peaked at $E_{\bar{\nu}}\sim$ 1.5 GeV). In
this case, the experiment informs that we have 0.44\% on neutrons
and 0.55\% of protons, and since our calculations were for free nucleons
we weight out results with these precentages depending on the channel.
Our results including all resonances and for $W_{\pi N}\lesssim2$
GeV are compared with the data reported in\cite{Bolog79} are shown
in Fig.(\ref{fig:Antinunocut}) , and as can be seen we get a consistent
description with that for the neutrino case.
We show results within the CMS with only the $\Delta$ and the $\Delta$
with the other resonances in the CMW, showing an appreciable difference
between the different approaches. Adding the energy dependent width
we get an improvement, but anyway we have consitence with the neutrino's
results.

\begin{center}
{\bf IV. CONCLUSIONS}
\end{center}

We have extended our dynamical model including resonances into the
first and second region, previously used to describe succesfully neutrino
and antineutrino scattering cross sections ANL data with the cuts
$W_{\pi N}\lesssim1.4,1.6$ GeV, to describe ANL and BNL data without
cuts. The model treats consitently the vertexes and propagators for
the spin-$\frac{3}{2}$ resonances from the point of view of contact
transformations of the spin-$\frac{3}{2}$ $\Psi_{\mu}$ field and
the role of the $\Psi_{0}$ component. In addition, we incorporate
the different pole resonant contributions and background resonant
and non-resonant ones, in a coherently sum to the amplitude. We have
added a global effective monopole FF to take into account the hadrons
size and possible more energetic resonances excitations not included
explicitly in the model for $W_{\pi N}\lesssim2.$GeV. The data to
compare, are that reanalyzed to eliminate flux normalization incertainties
in the ANL and BNL experiments \cite{wilki14,Rodriguez16}. 
In GENIE simulation to describe the mentioned data \cite{Rodriguez16},
single pion production is separated into resonant and non-resonant
terms, with interference between them neglected and interferences
between resonances neglected too in the calculation. The resonant
component is a modified version of the RS model \cite{Rein81}, where
the production and subsequent decay of 18 nucleon resonances with
invariant masses$W\le2$ GeV are considered. In GENIE, only 16 resonances
are included, based on the recommendation of the Particle Data Group
\cite{PDG06}. In this work they make the assumption that interactions
on deuterium can be treated as interactions on quasi-free nucleons
which are only loosely bound together, and so neglect FSI effects.
In GENIE, there are a number of systematic parameters which can be
varied to change the single pion production model. Resonant axial
mass ($M_{A}^{RES}$) Resonant normalization (RES norm) Non-resonant
normalization (DIS norm), and normalization of the axial form factor
($F_{A}(0)$). The total GENIE prediction is the incoherent sum of
the RES and DIS contributions, where interference terms have been
neglected. GENIE cannot describe all of the pion production channels
well for the reanalyzed datasets. For example, the data of the
 $\nu_{\mu}n\rightarrow\mu^{-}p\pi^{0},\nu_{\mu}n\rightarrow\mu^{-}n\pi^{+}$
channels are very similar, but there are large differences between
the nominal GENIE predictions for these channels. The non-resonant
component of the GENIE prediction, which contributes strongly to these
channels, appears to be too large. Nevertheless, within our model
these two channels are described properly. Finally, it can be seen
from Fig.(3) of Ref.\cite{Rodriguez16}, where neutrino energy distribution
is shown, that the nominal GENIE prediction fails to describe the
low-$Q^{2}$ data well for some channels.We also note that the GENIE
uncertainties are larger than the data suggests, and they may be reduced
by tuning the GENIE model to the ANL and BNL data. At difference,
within our model the low $Q^{2}$ distribution seem to be right. In
addition, we describe also and properly antineutrino cross section
without cuts in the data. In resume, its seem that keeping control
on the cross sections until the second resonance region ($W_{\pi N}\lesssim1.6$
GeV), enables a good description without cuts introducing the effect
of finite size of hadrons and more energetic resonances in an effective
way.

\section{Appendix}

The background contributions to the amplitude read 
\begin{eqnarray}
O_{\text{B}}^{\lambda}(p,p',q) & = & O_{\text{BN}}^{\lambda}(p,p',q)+O_{\text{BR}}^{\lambda}(p,p',q)\nonumber \\
\nonumber \\
\nonumber \\
O_{\text{BN}}^{\lambda}(p,p',q) & = & -i\frac{1}{2}\Bigg[F_{1}^{V}(Q^{2})\gamma^{\lambda}-i\frac{F_{2}^{V}(Q^{2})}{2m_{_{N}}}\sigma^{\lambda\nu}q_{\nu}-F^{A}(Q^{2})\gamma^{\lambda}\gamma_{_{5}}\Bigg]i\frac{\ps'+\qs+m_{_{N}}}{(p'+q)^{2}-m_{_{N}}}\nonumber \\
 & \times & \frac{g_{_{\pi NN}}}{2m_{_{N}}}\gamma_{_{5}}(\ps-\ps'-\qs)\sqrt{2}\mathcal{T}_{a}(m_{t},m_{t'})\nonumber \\
 & + & \frac{g_{_{\pi NN}}}{2m_{_{N}}}\gamma_{_{5}}(\ps-\ps'-\qs)i\frac{\ps-\qs+m_{_{N}}}{(p-q)^{2}-m_{_{N}}^{2}}\left(-i\frac{1}{2}\right)\Bigg[F_{1}^{V}(Q^{2})\gamma^{\lambda}-i\frac{F_{2}^{V}(Q^{2})}{2m_{_{N}}}\sigma^{\lambda\nu}q_{\nu}-F^{A}(Q^{2})\gamma^{\lambda}\gamma_{_{5}}\Bigg]\nonumber \\
 & \times & \sqrt{2}\mathcal{T}_{b}(m_{t},m_{t'})\nonumber \\
 & - & \frac{i}{(p-p')^{2}-m_{\pi}^{2}}iF_{1}^{V}(Q^{2})(2p-2p'-q)^{\lambda}\times\frac{g_{_{\pi NN}}}{2m_{_{N}}}\gamma_{_{5}}(\ps-\ps')\sqrt{2}\mathcal{T}_{c}(m_{t},m_{t'})\nonumber \\
 & + & \frac{g_{_{\pi NN}}}{2m_{_{N}}}F_{1}^{V}(Q^{2})\gamma_{_{5}}\gamma^{\lambda}\sqrt{2}\mathcal{T}_{d}(m_{t},m_{t'})\nonumber \\
 & + & i\frac{g_{_{\omega\pi V}}}{m_{\pi}}F_{1}^{V}(Q^{2})\epsilon^{\lambda\alpha\beta\delta}q_{_{\alpha}}(p-p')_{_{\beta}}i\frac{-g_{_{\delta\epsilon}}}{(p-p')^{2}-m_{\omega}^{2}}(-i)\frac{g_{_{\omega NN}}}{2}\Bigg[\gamma^{\epsilon}-i\frac{\kappa_{\omega}}{2m_{_{N}}}\sigma^{\epsilon\kappa}(p-p')_{\kappa}\Bigg]\nonumber \\
 & \times & \sqrt{2}\mathcal{T}_{e}(m_{t},m_{t'})\nonumber \\
 & + & f_{\rho\pi A}F^{A}(Q^{2})i\frac{-g^{\lambda\mu}}{(p-p')^{2}-m_{\rho}^{2}}(-i)\frac{g_{_{\rho NN}}}{2}\Bigg[\gamma_{\mu}-i\frac{\kappa_{\rho}}{2m_{_{N}}}\sigma_{\mu\kappa}(p-p')^{\kappa}\Bigg]\sqrt{2}\mathcal{T}_{f}(m_{t},m_{t'}),\label{eq:OBN}
\end{eqnarray}
\begin{eqnarray}
O_{\text{BR}}^{\lambda}(p,p',q) & =- & i\frac{1}{2}\Bigg[\frac{g_{1V}^{1440}}{(m_{_{1440}}+m_{_{N}})^{2}}(Q^{2}\gamma^{\lambda}+\qs q^{\lambda})-\frac{g_{2V}^{1440}}{(m_{_{1440}}+m_{_{N}})}i\sigma^{\lambda\nu}q_{\nu}-g_{1A}^{1440}\gamma^{\lambda}\gamma_{5}+\frac{g_{3A}^{1440}}{m_{_{N}}}q^{\lambda}\gamma_{5}\Bigg]\nonumber \\
 & \times & i\frac{\ps'+\qs+m_{R}}{(p'+q)^{2}-m_{1440}^{2}+i\Gamma_{1440}m_{1440}}(-)\frac{f_{_{1440\pi N}}}{m_{\pi}}\gamma_{5}(\ps-\ps'-\qs)\sqrt{2}\mathcal{T}_{g}^{1440}(m_{t},m_{t'})\nonumber \\
 & - & i\frac{1}{2}\gamma_{5}\Bigg[\frac{g_{1V}^{1535}}{(m_{_{1535}}+m_{_{N}})^{2}}(Q^{2}\gamma^{\lambda}+\qs q^{\lambda})-\frac{g_{2V}^{1535}}{(m_{_{1535}}+m_{_{N}})}i\sigma^{\lambda\nu}q_{\nu}-g_{1A}^{1535}\gamma^{\lambda}\gamma_{5}+\frac{g_{3A}^{1535}}{m_{_{N}}}q^{\lambda}\gamma_{5}\Bigg]\nonumber \\
 & \times & i\frac{\ps'+\qs+m_{1535}}{(p'+q)^{2}-m_{1535}^{2}+i\Gamma_{1535}m_{1535}}(-)\frac{f_{_{1535\pi N}}}{m_{\pi}}(\ps-\ps'-\qs)\sqrt{2}\mathcal{T}_{g}^{1535}(m_{t},m_{t'})\nonumber \\
 & + & (-\overline{){W_{\lambda\alpha}^{WN\Delta}(p,p',-q)}}iG_{\Delta}^{\alpha\beta}(p'+q)(-)\frac{f_{_{\Delta\pi N}}}{m_{\pi}}(p-p'-q)_{\beta}\sqrt{2}\mathcal{T}_{g}^{\Delta}(m_{t},m_{t'})\nonumber \\
 & + & (-\overline{){\frac{1}{2}W_{\lambda\alpha}^{WN1520}(p,p',-q)}}iG_{1520}^{\alpha\beta}(p'+q)(-)\frac{f_{_{1520\pi N}}}{m_{\pi}}\gamma_{5}(p-p'-q)_{\beta}\sqrt{2}\mathcal{T}_{g}^{1520}(m_{t},m_{t'}),\label{OBR}
\end{eqnarray}
where $\frac{1}{2}$-resonances the are put before the $\frac{3}{2}$-ones.
 $G_{R}^{\alpha\beta}$ togheter $W^{WNR}=\left(V^{WNR}+A^{WNR}\right)$are
the spin-$\frac{3}{2}$ resonance propagators and $WN\rightarrow R$
vertexes respectively, both defined in Ref.\cite{Tamayo22} while
$\overline{W}=\gamma_{0}W^{\dagger}\gamma_{0}$ . $V^{WNR}$ is the
vector vertex as in pion-photo($Q^{2}=0)$\cite{Mariano07} and electroproduction
applying CVC within the Sachs parametrization, and $A^{WNR}$ the
axial contribution compatible with $V_{\nu\mu}^{WN\Delta}$ \cite{Barbero08}(it
could be, in principle, obtained by using $-V_{\nu\mu}^{WN\Delta}\gamma_{_{5}}$).
\\
The corresponding pole contributions coming from the resonances are

\begin{eqnarray}
\mathcal{O}_{R}^{\lambda} & = & \frac{f_{_{1440\pi N}}}{m_{\pi}}\gamma_{5}(\ps-\ps'-\qs)i\frac{\ps-\qs+m_{R}}{(p-q)^{2}-m_{1440}^{2}+i\Gamma_{1440}m_{1440}}\nonumber \\
 & \times & (-i)\frac{1}{2}\Bigg[\frac{g_{1V}^{1440}}{(m_{_{1440}}+m_{_{N}})^{2}}(Q^{2}\gamma^{\lambda}+\qs q^{\lambda})-\frac{g_{2V}^{1440}}{(m_{_{1440}}+m_{_{N}})}i\sigma^{\lambda\nu}q_{\nu}-g_{1A}^{1440}\gamma^{\lambda}\gamma_{5}+\frac{g_{3A}^{1440}}{m_{_{N}}}q^{\lambda}\gamma_{5}\Bigg]\sqrt{2}\mathcal{T}_{h}^{1440}(m_{t},m_{t'})\nonumber \\
 & + & (-)\frac{f_{_{1535\pi N}}}{m_{\pi}}(\ps-\ps'-\qs)i\frac{\ps-\qs+m_{1535}}{(p-q)^{2}-m_{1535}^{2}+i\Gamma_{1535}m_{1535}}\nonumber \\
 & \times & (-i)\frac{1}{2}\Bigg[\frac{g_{1V}^{1535}}{(m_{_{1535}}+m_{_{N}})^{2}}(Q^{2}\gamma^{\lambda}+\qs q^{\lambda})-\frac{g_{2V}^{1535}}{(m_{_{1535}}+m_{_{N}})}i\sigma^{\lambda\nu}q_{\nu}-g_{1A}^{1535}\gamma^{\lambda}\gamma_{5}+\frac{g_{3A}^{1535}}{m_{_{N}}}q^{\lambda}\gamma_{5}\Bigg]\gamma_{5}\mathcal{T}_{h}^{1335}(m_{t},m_{t'})\nonumber \\
 & + & (-)\frac{f_{_{\Delta\pi N}}}{m_{\pi}}(p-p'-q)_{\alpha}iG_{\Delta}^{\alpha\beta}(p-q)W_{\beta\lambda}^{WN\Delta}(p,p',q)\sqrt{2}\mathcal{T}_{g}^{\Delta}(m_{t},m_{t'})\nonumber \\
 & + & (-)\frac{f_{_{1520\pi N}}}{m_{\pi}}\gamma_{5}(p-p'-q)_{\alpha}iG_{\Delta}^{\alpha\beta}(p-q)W_{\beta\lambda}^{WN1520}(p,p',q)\sqrt{2}\mathcal{T}_{g}^{1520}(m_{t},m_{t'}),\label{eq:OR}
\end{eqnarray}
where all  the vector and axial weak FF present in Eqs.(\ref{eq:OBN}-\ref{eq:OR}) 
are defined in Ref.\cite{Tamayo22}. Note that the $\frac{1}{2}$
factor in the weak vertex of the spin-$\frac{1}{2}$ resonances comes
from the isovector part of the charge operator $\frac{\tau_{3}}{2}$
dragged from the CVC hypotesis.

\section{Acknowledgments}

A. Mariano belong to CONICET and UNLP, D.F. Tamayo Agudelo and D.E.
Jaramillo Arango to UdeA.


\begin{thebibliography}{10}
\bibitem[1]{Eberly15}1. B.Eberly et al.,Phys.Rev.D92(9),092008(2015). 

\bibitem[2]{Le15}T.Le et al.,Phys.Lett.B749,130(2015).

\bibitem[3]{Are11} A. Aguilar-Arevalo et al., Phys. Rev. D 83, 052007
(2011). 

\bibitem[4]{Sob15} J.T.Sobczyk,J.Z\.{ }muda,Phys.Rev.C91(4),045501(2015).

\bibitem[5]{Ulrich15}5. U. Mosel, Phys. Rev. C 91(6), 065501 (2015). 

\bibitem[6]{Rad82} G. M. Radecky, et. al, Phys. Rev. D \textbf{25,}
1161 (1982). 

\bibitem[7]{Kitagi86}T. Kitagaki,et al., Phys. Rev. D \textbf{34
}, 2554 (1986). 

\bibitem[8]{Nieves07}E. Hernandez, J. Nieves, M. Valverde, Phys.
Rev. D 76, 033005 (2007). 

\bibitem[9]{Leitner09} T. Leitner, O. Buss, L. Alvarez-Ruso, and
U. Mosel,\textsf{\large{} }\textit{\emph{Phys. Rev. C 79,(2009).}}

\bibitem[10]{Nieves10}E. Hernandez, J. Nieves, M. Valverde, M. Vicente,
Vacas. Phys. Rev. D 81, 085046 (2010).

\bibitem[11]{Tamayo22}D.F. Tamayo Agudelo, A. Mariano, D.E. Jaramillo
Arango, Phys. Rev. D 105, (2022) 033008.

\bibitem[12]{Ahn03} M. Ahn et al., Phys. Rev. Lett. 90, 041801 (2003). 

\bibitem[13]{Abe13} K. Abe et al., Phys. Rev. D 88, 032002 (2013). 

\bibitem[14]{Lala13}O. Lalakulich, U. Mosel, Phys. Rev. C 87, 014602
(2013). 

\bibitem[15]{Graczyk09}K.Graczyk,D.Kielczewska,P.Przewlocki,J.Sobczyk,Phys.Rev.
D 80, 093001 (2009). doi:10.1103/PhysRevD.80.093001 

\bibitem[16]{Graczyk14}K.M. Graczyk, J. Zmuda, J.T. Sobczyk, Phys.
Rev. D 90, 093001 (2014). doi:10.1103/PhysRevD.90.093001 

\bibitem[17]{wilki14}C. Wilkinson, P. Rodrigues, S. Cartwright, L.
Thompson, K. McFarland, Phys. Rev. D 90(11), 112017 (2014). doi:10.1103/
Phy

\bibitem[18]{Rodriguez16}Philip Rodrigues, Callum Wilkinson,and ,Kevin
McFarland, Eur. Phys. J. C (2016) 76:474. 

\bibitem[19]{Genie10}C. Andreopoulos, A. Bell, D. Bhattacharya, F.
Cavanna, J. Dobson et al., Nucl. Instrum. Methods A A614,87 (2010).

\bibitem[20]{Rein81}D. Rein, L.M. Sehgal, Ann. Phys. 133(1), 79 (1981). 

\bibitem[21]{Adanson2016}NOvA, P. Adamson et al., First measurement
of muon-neutrino disappearance in NOvA. Phys. Rev. D 93(5), 051104
(2016). doi:10.1103/PhysRevD.93.051104.

\bibitem[22]{Kirbach2002}M. Kirchbach and D. Ahluwalia, Phys.Lett.\textbf{B},
(2002) 529124.

\bibitem[23]{Badagnani12}D. Badagnani, A. Mariano, and C. Barbero,
J. Phys. G: Nucl. Part. Phys. \textbf{39 }(2012) 035005. 

\bibitem[24]{Badga17}D. Badagnani, A. Mariano, and C. Barbero, J.
Phys. G: Nucl. Part. Phys. 44, 025001 (2017). 

\bibitem[25]{Mariano11}A. Mariano, C. Barbero and
G. L\'opez Castro, Nuclear Physics A 849 (2011) 218. 

\bibitem[26]{Barbero08}\textsf{\large{} }C. Barbero, A. Mariano,
G. Lopez Castro. Physics letters B 664 (2008) 70-77.\textsf{\large{} }{\large \par}


\bibitem[27]{Barbero15}C. Barbero, A. Mariano, J. Phys. G: Nucl.
Part. Phys. 42 (2015) 105104 .

\bibitem[28]{Mariano07}\textsf{\large{} }A. Mariano. Phys. lett.
B (2007) 253; A. Mariano, J. Phys. G (2007) 1627.

\bibitem[29]{Amiri92}M. el Amiri, J. Pestieau and G. L\'pez Castro,
Nucl. Phys. A 543( 1992) 673.

\bibitem[30]{Mariano01}G. Lopez Castro, A. Mariano, Nucl. Phys. A
697 (2001)440.

\bibitem[31]{Melde09}T. Melde,L. Canton,and W. Plessas, PRL 102,
132002 (2009).

\bibitem[32]{Mosel98}T. Feuster, and U. Mosel, Phys. Rev. C 58,(1998)457.T.
Sato and T.-S. H. Lee, PHYSICAL REVIEW \textbf{C54}, (1996)2660.

\bibitem[33]{Sato96}T. Sato and T.-S. H. Lee, Phys. Rev. C 54, (1996)
2660.

\bibitem[34]{Pearce91}B.C. Pearce and B.K. Jennings, NuclearPhysicsA528
(1991) 655-675.

\bibitem[35]{footnote}At first, it is possible to manipule the FF proposed in that references
to get forms very similar to Eq.(\ref{eq:FF4mod-1}).

\bibitem[36]{Barbero14} C. Barbero, A. Mariano, G. Lopez Castro.
Physics Letters B 728 (2014) 282-287. 

\bibitem[37]{Lakakulich06}O. Lalakulich, E. A. Paschos, and G. Piranishvili,
Phys. Rev. D 74,(2006), 014009 .

\bibitem[38]{Bolog79}T. Bolognese, J.P. engel, J.L Guyonnet and J.L.
Riester, Phys. Lett. 81B (1979),393.

\bibitem[39]{PDG06}Review of Particle Physics,W.-M. Yao, et al, J.
Phys. G 33 (2006) 1. 
\end{thebibliography}
\end{document}